\documentclass[12pt]{article}
\usepackage{amssymb}
\usepackage{amsfonts}
\usepackage{amsmath,amsthm}
\usepackage{graphics,epsfig}
\usepackage{subfigure}

\oddsidemargin 10mm \numberwithin{equation}{section}
\def\be{\begin{equation}}
\def\ee{\end{equation}}
\begin{document}
\begin{center} {{\bf {Thermodynamic phase transition for Quintessence Dyonic Anti de Sitter  Black Holes}}\\
 \vskip 0.5 cm
  {{ Hossein Ghaffarnejad \footnote{E-mail: hghafarnejad@semnan.ac.ir
 } }{ Emad Yaraie \footnote{E-mail: eyaraie@semnan.ac.ir
 } }{ and Mohammad Farsam \footnote{E-mail: mhdfarsam@semnan.ac.ir
 } }}\vskip 0.1 cm \textit{Faculty of Physics, Semnan
University, P.C. 35131-19111, Semnan, Iran }}
\end{center}
\begin{abstract}
We study thermodynamic of a dyonic AdS black hole surrounded by
quintessence dark energy where negative cosmological constant of
AdS space behaves as pressure of the black hole. We choose grand
canonical ensemble of the black hole where its   magnetic charge
$Q_M$  and electric potential $\Phi_E$ are hold as constant. Our
goal  in this work is study physical effects of the magnetic
charge and electric potential on the thermodynamic phase
transition of the black hole in presence of quintessence dark
energy. When barotropic index of the quintessence is
$\omega=-\frac{7}{9}$ we obtained that compressibility factor of
the black hole reduces to $Z_c=\frac{3}{8}$ which corresponds to
the Van der Waals fluid. We obtained analogy between the
small/large black hole phase transition and liquid/gas phase
transition of the Van der Waals fluid. Numerical calculations
predict that the black hole may born $plasma$ phase which is
fourth different state of the matter which does not appear in the
Van der Waals fluid.
\end{abstract}
\section{Introduction}
~~Thermodynamic aspects of black holes have  challenge when the
argument of thermodynamic phase transition is under consideration.
This is because of association of the phase transition to the
 thermodynamic variables for instance entropy, volume  and etc. One
of most controversial thermodynamic variables is volume of black
holes undoubtedly. Consider a flat space-time including a black
hole and an observer  located at distances far from the black hole
which can not observe small region of space-time hidden behind the
black hole horizon. In fact this region which is equal with
geometric volume of the black hole doesn't have an straightforward
definition due to use of different gravity models. It is well
known that nonzero value of the vacuum energy density of the de
Sitter space is relating to the cosmological constant. In the
latter case a typical definition for the volume is there: That is
related to total energy of the gravitational system but not to the
black hole mass amount which is stand as hidden from point of view
of an observer far from the black hole (ADM mass [1,2,3]). This
means that the hidden energy from point of view of observer would
be part of total energy which is equal to the ADM mass located on
hidden volume. In the thermodynamic of the black holes the
observed mass by an observer far from the black hole is considered
to be the black hole thermal energy while the cosmological
constant behaves as a fixed parameter. For an anti de-Sitter space
the cosmological constant has negative values arising from an
energy-momentum tensor which reads the vacuum equation of state
$P=-\rho=-\frac{\Lambda}{8\pi G_N}.$ Regarding the latter concept
one can write first law of black holes thermodynamic as follows.
\begin{equation}
dM=TdS+VdP~~\Rightarrow~~dU=TdS-PdV
\end{equation}
where $U=M-PV$ and $H=U+PV$ are internal energy and enthalpy of
the black holes respectively. This  shows the black hole mass $M$
can interpreted as enthalpy. In this view the cosmological
constant is related to pressure of the AdS space-time [4]. Since
the cosmological constant corresponds to space-time pressure thus
its conjugate variable must have volume dimension. This is called
thermodynamic volume for which we have $V=(\frac{\partial
M}{\partial P})_{S}$ [5]. In general, this definition of the black
hole volume is different with the geometric volume which we
pointed previously. This contains some universal properties which
they satisfy the condition of reverse iso-perimetric inequality
[6]. Regarding the above definition for the black hole
thermodynamic volume and the pressure, one can study critical
behavior of the black holes in the P-V frame and in the extended
phase space [7] (see also [8] for charged and [9] for rotating
black holes). For a dyonic black hole, authors of ref. [10]
studied magnetic charge effects on T-V critically of the black
holes via the holographic approach. They observed that in constant
electric potential and constant magnetic charge, the phase diagram
of ensemble of dyonic black hole is similar to well known phase
transition of Van der Waals fluid done in presence of its internal
chemical potential. They obtained ferromagnetic like behavior of
boundary theory of the dyonic black hole when the external
magnetic field vanishes. Authors of the work [11] studied phase
structure of the quintessence Reissner-Nordstr\"{o}m-AdS black
hole with the nonlocal observables such as holographic
entanglement entropy and two point correlation functions. Results
of their work show that, as the case of the thermal entropy, both
the observables exhibit the similar Van der Waals-like phase
transition. They were check the equal area law  and critical
exponent of the heat capacity for the first and the second order
phase transition respectively. Also they discussed the effect of
the parameter of state equation on the phase structure of the
nonlocal observables. Hartnoll et al studied AdS/CFT
superconductor properties of a planar AdS Schwarzschild black hole
which for temperatures less than the critical temperature, reduces
to  a charged condensate state. This is formed via a second order
phase transition for which the conductivity reaches to infinite
value [12,13, 14,15,16]. The study of the  phase transition of the
dyonic black hole surrounded by dark energy must be very complex
and attractive because of dependence to the possibility existence
of the dark energy. In fact There is a wide range of modern
cosmological observations which confirm that our universe expands
with positive acceleration. This leads us to infer that instead of
the Newton`s gravity force there should exist other forces which
is not still observed [17,18,19]. Dark energy is one of the most
important candidates to explain this situation, which must
occupied seventy percent of energy of our universe. Among diverse
dark energy models, `quintessence` as a canonical scalar field can
be a good model to explain acceleration of the universe
[20,21,22]. There are many works where various gravitational
models are used to study thermodynamic aspect of black holes  (see
for instance [23-26]).
\\
In this work we study thermodynamics of the AdS dyonic black hole
which is surrounded by the quintessence dark energy.
 We calculate
its equation of state on the event horizon hypersurface where the
cosmological constant behaves as the black hole pressure. Then for
different values of barotropic index of the quintessence dark
energy $\omega$ we obtained parametric forms of critical points
$\{T_c,V_c,P_c\}$ defined in terms of the constant electric
potential and magnetic charge and $\omega$. Then we set
$\omega=-\frac{7}{9}$ to obtain numerical values of the critical
points. Because for this particular value of the quintessence
barotropic index the black hole compressibility factor reaches to
one which is obtained for the Van der Waals fluid as
$Z=\frac{3}{8}.$ This restriction helps us to improve statement
which  extract from the work.
 The organization of the work is as follows.\\ In section 2 we review
briefly phase transition of a Van der Waals fluid.
 In section 3 we define a dyonic AdS black hole in presence of quintessence dark
energy counterpart. Then we obtain its equation of state by
solving equation of the event horizon. Enthalpy, heat capacity at
constant pressure and Gibbs free energy are other thermodynamic
variables of the black hole which we calculated in this section.
At last we plot variations of the these variables versus the
volume and the temperature. Also we obtained all possible
numerical values for the critical points at zero electric
potential $\Phi_E=0$ versus different values of the magnetic
charge $q_M.$ Section 4 denotes to results of the this work and
outlooks.
\section{Phase transition of Van der Waals fluid}
~~ Equation of state for an ideal gas has simple form as
$PV=Nk_BT$ in which for a constant temperature (isotherm curves)
the pressure decreases absolutely by increasing the volume. While
this is not ready for a real `Van der Waals` fluid. In the latter
case, the equation of state  given by the following form behaves
complex [27].
\begin{equation}
P(T,V)=\frac{Nk_BT}{V-Nb}-\frac{aN^2}{V^2},
\end{equation}
 where $N$ is number of particles in the fluid and the constant $b$ is intended to correct for the volume occupied by the molecules and the
  term $\frac{aN^2}{V^2}$ is a correction that accounts for the intermolecular forces
 of attraction. In fact, these constants are evaluated by noting that the critical isotherm passes through a
  point of inflection at the critical point and that the slope is zero at this point.
 The critical point $\{T_c,V_c,P_c\}$ is determined by solving
 \begin{equation}
\big(\frac{\partial P}{\partial V}\big)_T=\big(\frac{\partial^2
P}{\partial V^2}\big)_T=0
\end{equation}
which read
\begin{equation}V_c=3bN,~~~P_c=\frac{a}{27b^2},~~~k_BT_c=\frac{8a}{27b}\end{equation} and one can calculate compressibility dimensionless factor $Z_c$
as follows.
\begin{equation} Z_c=\frac{P_cV_c}{Nk_BT_c}=\frac{3}{8}.\end{equation}
Substituting
 the dimensionless thermodynamic variables $p=\frac{P}{P_c},~v=\frac{V}{V_c},~t=\frac{T}{T_c}$ one can infer the equation of state
 (2.1) reads to a dimensionless form as follows (see ref. [27]).
\begin{equation}p(v,t)=\frac{8t}{3v-1}-\frac{3}{v^2}.\end{equation}
Gibbs free energy difference for a Van der Waals fluid is given by
\begin{equation}dG=-SdT+VdP\end{equation} where $S$
 is entropy. For a constant temperature the first term in right hand side of the above equation is eliminated and so we
obtain \begin{equation}G=\int VdP=PV-\int P(V)dV\end{equation}
which for the dimensionless equation of state (2.5) reads
\begin{equation} g(t,v)=\frac{8t}{3}\bigg[\frac{3v}{3v-1}-\ln(3v-1)\bigg]-\frac{6}{v}.\end{equation}
Holding the temperature as constant, we plot diagrams for the
equation of state (2.5) and for the Gibbs free energy (2.8) in
figures 1-c and 1-b respectively. Here $t=0.5<t_c,$ $t=t_c=1$ and
$t=1.5>t_c$ correspond to dash, solid and dash-dot lines
respectively.  (1-c) shows for $t<t_c$ there is a phase transition
for the fluid which coexistence of two phase occurs at the minimum
pressure. This point has minimum Gibbs free energy which is shown
in figure (1-b). Figure (1-a) shows variation of the Gibbs free
energy versus the temperature. It shows at all, the Van der Waals
fluid has different phases which they cross each other at the
phase transition point. This point corresponds to coexistence
regime of both the gas and the fluid phases of the Van der Waals
matter which occurs at small scale of thermodynamic volume.
\begin{figure}[ht] \centering \subfigure[{}]{\label{1}
\includegraphics[width=.3\textwidth]{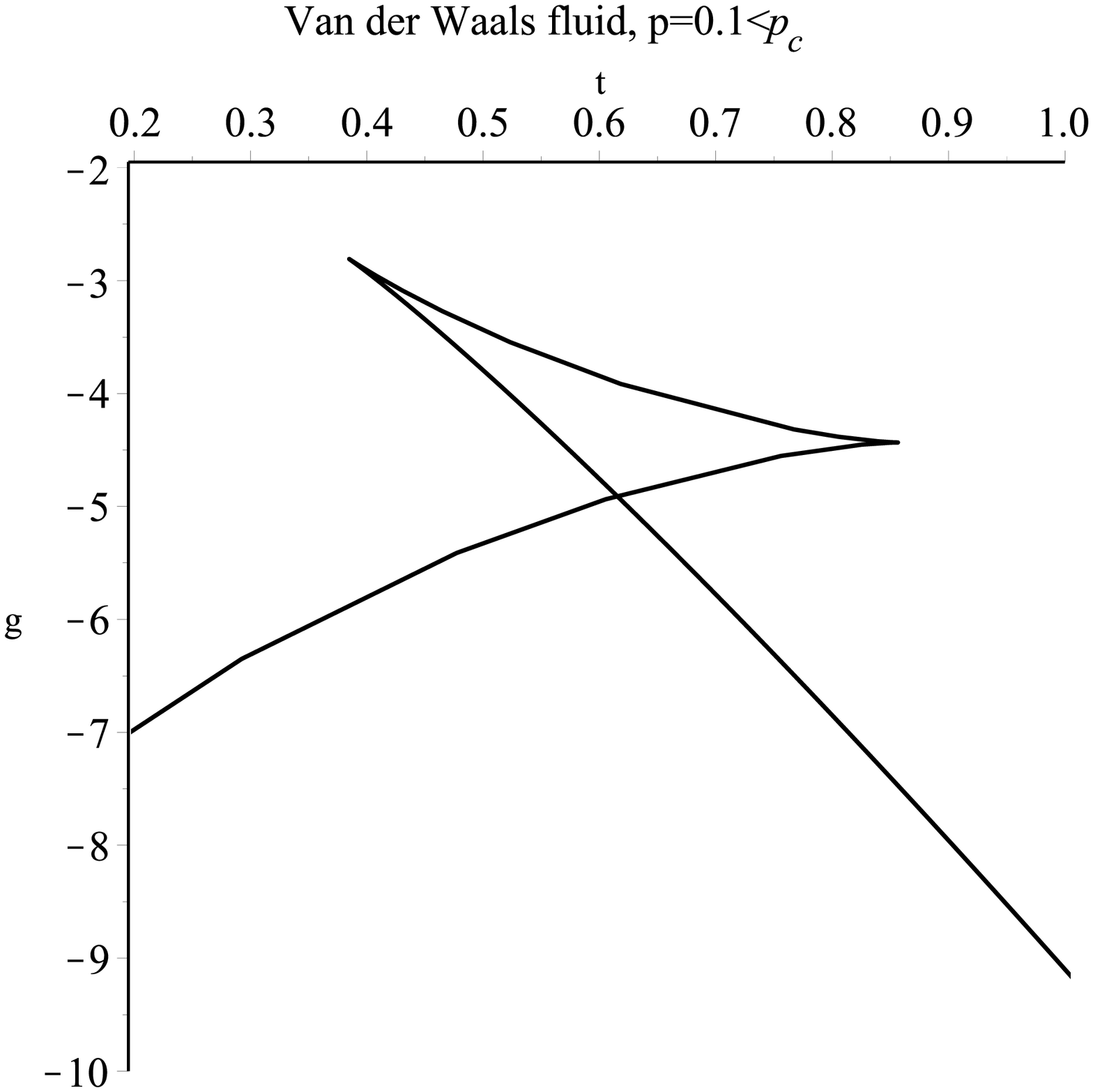}}
\hspace{1mm} \subfigure[{}]{\label{1}
\includegraphics[width=.3\textwidth]{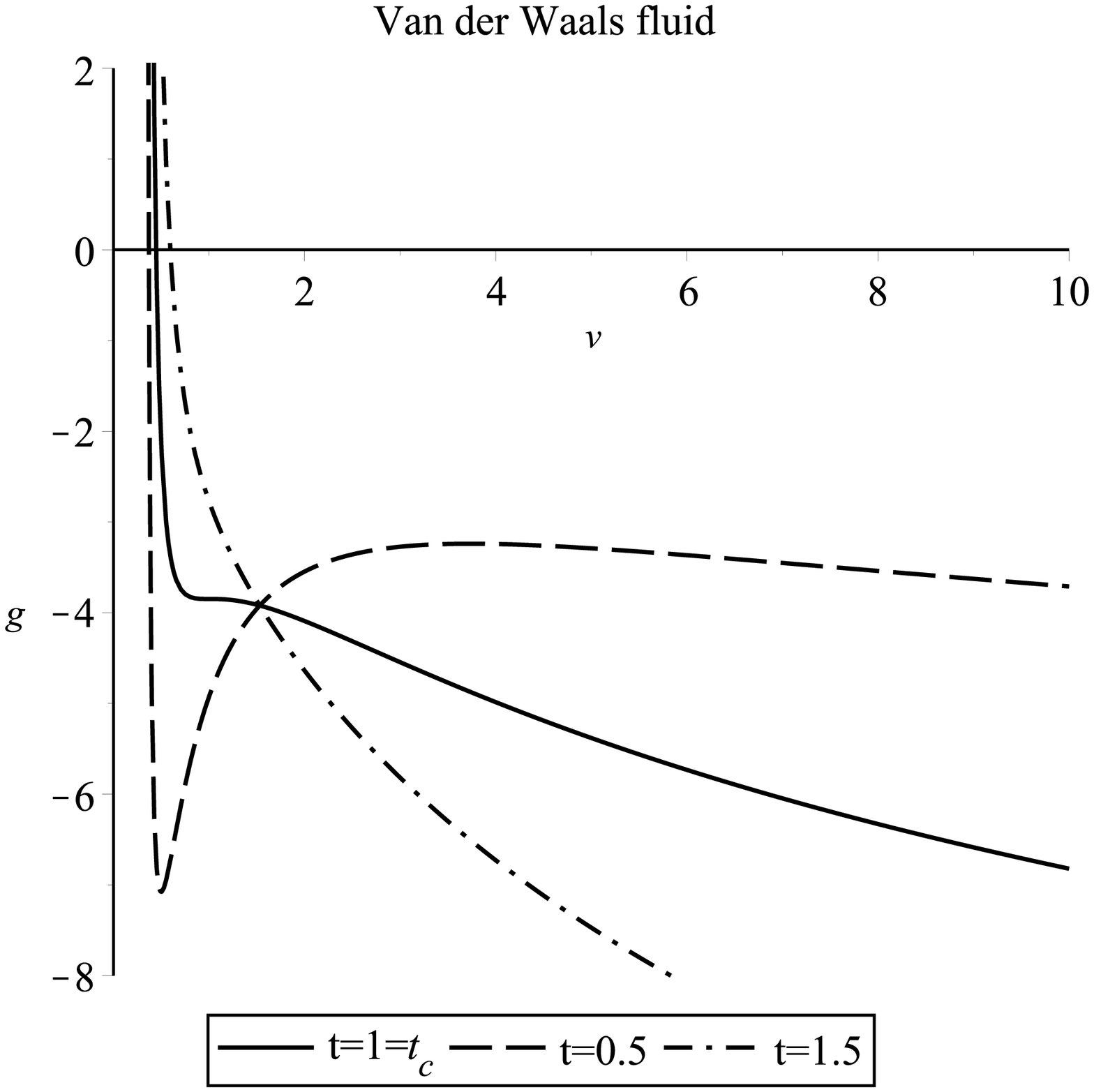}}
\hspace{1mm} \subfigure[{}]{\label{1}
\includegraphics[width=.3\textwidth]{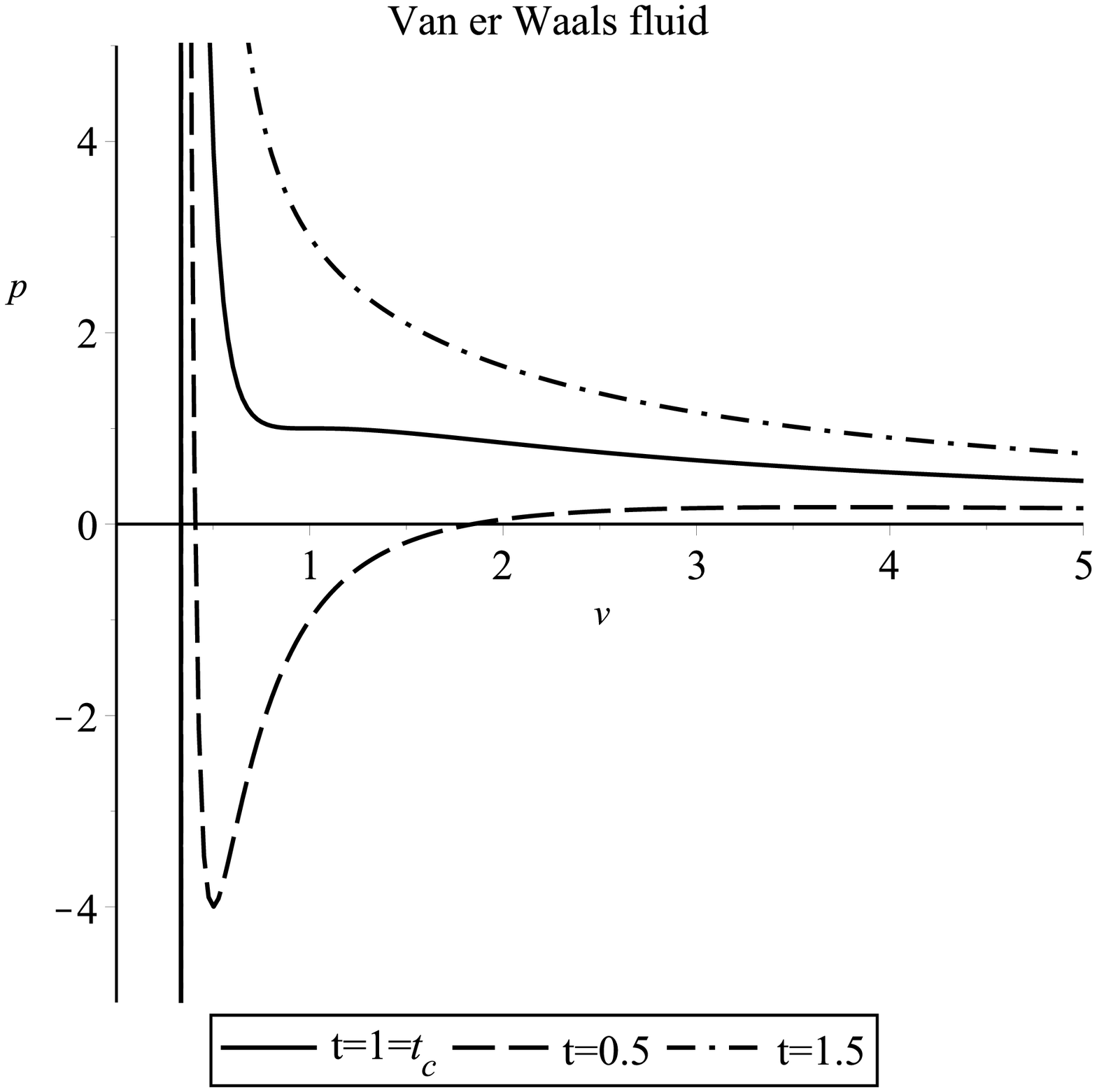}}
\caption{Diagrams of Gibbs free energy and pressure of a Van der
Waals fluid are plotted versus the temperature $g-t$ at (a) and
against the volume $g-v$ and $p-v$ at (b) and (c) respectively by
setting $p_c=1.$} \label{l}
\end{figure}
We obtained by numerical calculations that for $p\geq p_c=1$ the
diagrams of $p-v$, $g-t$ and or $g-v$ have not a minimum point in
phases space (not shown) and so the phase transition dose not
appeared in the latter case
.\\
Our aim in this work is to study thermodynamic phase transition of
a quintessence AdS dyonic black hole and obtain situations for
which the black hole behaves thermodynamically as a Van der Waals
fluid.
\section{ Quintessence Dyonic AdS black hole thermodynamic} According to the work presented by Kiselev in [23] for metric of a
spherically symmetric static Reissner Nordstr\"{o}m black hole
surrounded by quintessence dark energy we use an extension of his
metric solution where an extra magnetic charge $Q_M$ is added and
so it is called as dyonic AdS quintessence black hole with the
following metric form.
\begin{equation}
ds^2=f(r)dt^2-\frac{dr^2}{f(r)}-r^2(d\theta^2+sin^2\theta
d\phi^2),
\end{equation}
in which for quintessence regimes $-1<\omega<-\frac{1}{3}$ we have
\begin{equation}
f(r)=1-\frac{2M}{r}+\frac{Q_M^2}{r^2}+\frac{Q_E^2}{r^2}-\bigg(\frac{r_q}{r}\bigg)^{1+3\omega}-\frac{\Lambda r^2}{3},\\
\end{equation}
where $r_q$  is the dimensional normalization constant and the
negative  cosmological constant
 $\Lambda<0$ of the AdS space behaves as thermodynamic pressure of the black hole. Quintessence counterpart of the above metric solution
 $(r_q/r)^{1+3\omega}$ comes from leading order terms of a convergent
 series solution  which is obtained originally by the Kiselev by applying the condition of linearity and additivity of the Einstein
 metric equation [23]. \\
As we know the phase structure depends on the statistical ensemble
which we choose to study the phase transition. There is no new
interesting results if the black hole phase transition is studied
by particular ensemble in which both of the electric and the
magnetic charges are hold as constant.  In the contrary if we hold
magnetic charge  together with the electric potential as constant
parameters then there is obtained possibly new physics from study
of the phase transition.
 Former case is a canonical ensemble where a
small/large black hole phase transition dose not appeared in
absence of the quintessence while the latter setting become grand
(mixed) canonical ensemble for which a small/large black hole
phase transition occurs [10, 28]. However the former situation is
similar to case where the electric charge is shifted by $Q_E^2\to
Q_E^2+Q_M^2$ while in the latter case the first law of
thermodynamics is described in extended phase space (see for
instance [29]).
 To approach our goal in this work we extend the first
 law of the black hole thermodynamic given by the equation (2.12)
 in ref. [30] as follows.

\begin{equation}
dM=TdS+\Phi_EdQ_E+\Phi_MdQ_M+VdP+Ad r_q
\end{equation}
where $A$ is conjugate variable for the normalization factor $r_q$
of the quintessence dark energy. It is defined by
\begin{equation} A=\bigg(\frac{\partial M}{\partial r_q}\bigg)_{S,Q_{E},Q_{M},P}.\end{equation}
 The Smarr relation is obtained from the mass function of the black hole because the black hole mass is equal to
  the enthalpy. Applying the black hole metric (3.2) and
  solving the horizon equation $f(r_h)=0$ we obtain mass of the AdS quintessence dyonic black hole versus the thermodynamics variables
  as follows.
 \begin{equation}M=\frac{r_h}{2}+\frac{Q_M^2}{2r_h}+\frac{Q_E^2}{2r_h}-\frac{r_q^{1+3\omega}}{2r_h^{3\omega}}+\frac{4\pi r_h^3}{3}P\end{equation}
where we substitute pressure $P$ of the AdS space with
corresponding radius $r_{AdS}$ as follows.
\begin{equation}P=-\frac{\Lambda}{8\pi},~~~~r_{AdS}=\sqrt{-\frac{3}{\Lambda}}.\end{equation}
Using (3.3) one can calculate thermodynamic volume of the
quintessence dyonic AdS black hole as
\begin{equation} V=\bigg(\frac{\partial M}{\partial P}\bigg)_{A,Q_{E},Q_{M},r_q}=\frac{4\pi r_h^3}{3}\end{equation}
which its form is similar to the geometric volume of the horizon
with surface era $4\pi r_h^2.$ We should notice that the latter
statement is not correct for metric of an arbitrary curved space
time and in general the black hole thermodynamic volume is differ
with its geometric one. From (3.5) we obtain
\begin{equation} \Phi_E=\bigg(\frac{\partial M}{\partial Q_E}\bigg)_{S,P,Q_{M},r_q}=\frac{Q_E}{r_h}\end{equation}
 which is called as the electric potential of the black hole on the horizon.
 Substituting (3.5) and calculating (3.4) we obtain
 \begin{equation}A=-\frac{(1+3\omega)}{2}\bigg(\frac{r_q}{r_h}\bigg)^{2\omega}.\end{equation}
 The Hawking temperature of the above black hole can be derived as follows.
 \begin{equation}T=\frac{f^{\prime}(r)}{4\pi}_{|_{r=r_h}}=\frac{1}{4\pi}\bigg(\frac{1}{r_h}-\frac{Q_M^2}{r_h^3}-\frac{Q_E^2}{r_h^3}+\frac{3\omega
 r_q^{1+3\omega} }{r^{3\omega+2}_h}+8\pi r_h P
 \bigg)\end{equation} where we substitute (3.5) to remove $M$. Since the black hole mass $M$ is interpreted as enthalpy as $H=M$
 thus one can infer that the equation (3.5)
  reads to the following form.
  \begin{equation}H=U+PV\end{equation} where
   \begin{equation} U=\frac{r_h}{2}+\frac{Q_M^2}{2r_h}+\frac{Q_E^2}{2r_h}-\frac{r_q^{1+3\omega}}{2r_h^{3\omega}}\end{equation}
   is internal energy. To study location of critical point on the P-V plan it is appropriate to use dimensionless forms of the thermodynamic
   functions as follows.
      \begin{equation}p=8\pi r_q^2P,~~~t=4\pi r_qT,~~~v=\frac{r_h}{r_q},~~~ q_M=\frac{Q_M}{r_q}\end{equation} Substituting
 the above definitions and (3.8) into the equation (3.10) we obtain dimensionless equation of state for
 the quintessence dyonic AdS black hole such that
\begin{equation}
p(v,t)=\frac{t}{v}-\frac{(1-\Phi_E^2)}{v^2}+\frac{q_M^2}{v^4}-
\frac{3\omega}{v^{3(1+\omega)}}.
\end{equation}
By applying the above dimensionless parameters the  enthalpy
(3.11) reads
\begin{equation}m=(1+\Phi_E^2)v+\frac{q_M^2}{v}-\frac{1}{v^{3\omega}}+\frac{pv^3}{3}\end{equation}
 in which
\begin{equation}m=\frac{M}{2r_q}\end{equation}
is assumed to be dimensionless mass of the black hole. Solving the
equations
\begin{equation} \bigg(\frac{\partial p}{\partial
v}\bigg)_{other~parameters}=\bigg(\frac{\partial^2 p}{\partial
v^2}\bigg)_{other~parameters}=0
\end{equation} we can obtain parametric form of the
critical point $\{p_c,t_c,v_c\}$ which satisfy the following
equations.
\begin{equation}
2(1-\Phi_E^2)v_c^{3\omega+1}-12q_M^2v_c^{3\omega-1}+9\omega(1+\omega)(2+3\omega)=0
\end{equation}
\begin{equation}
t_c=\frac{2(1-\Phi_E^2)}{v_c}-\frac{4q_M^2}{v_c^3}+\frac{9\omega(1+\omega)}{v_c^{2+3\omega}}\end{equation}
\begin{equation}
p_c=\frac{(1-\Phi_E^2)}{v_c^2}-\frac{3q_M^2}{v_c^4}+\frac{3\omega(2+3\omega)}{v_c^{3(1+\omega)}}
.\end{equation}   The above algebraic equations have several roots
due to different values of the black hole parameters
$\{\omega,q_M,\Phi_E\}$ which are into account where
$-1<\omega<-\frac{1}{3}$ corresponds to the quintessence dark
energy regime.  To compare compressibility factor of this black
hole with which one obtained for the Van de Waals fluid as
$Z=\frac{3}{8}$ given by the equation (2.4), it is appropriate to
substitute $1-\Phi^2_E$ from (3.18) into the equations (3.19) and
(3.20). In the latter case (3.19) and (3.20) reduce to the
following forms respectively. \begin{equation}
t_c=\frac{8q_M^2}{v_c^3}-\frac{9\omega(1+\omega)^2}{v_c^{2+3\omega}}\end{equation}
and
\begin{equation}Z=\frac{p_cv_c}{t_c}=\frac{3}{8}-\frac{3\omega(1+3\omega)(7+9\omega)}{8t_cv_c^{2+3\omega}}.
\end{equation}
The above relation shows that for a quintessence regime of the
dark energy where $-1<\omega<-\frac{1}{3}$ the compressibility of
the dyonic AdS black hole behaves as a Van der Waals fluid if we
set
\begin{equation}\omega_{Van}=-\frac{7}{9}.\end{equation}  Substituting (3.23) into the equations
(3.14), (3.18), (3.21) and (3.22) we will have respectively
\begin{equation}p=\frac{t}{v}-\frac{(1-\Phi^2_E)}{v^2}+\frac{q_M^2}{v^4}+\frac{7}{3}\frac{1}{v^{\frac{2}{3}}}\end{equation}
\begin{equation}1-\Phi_E^2=\frac{6q_M^2}{v_c^2}-\frac{7}{27}v_c^\frac{4}{3},\end{equation}
\begin{equation}t_c=\frac{8q_M^2}{v_c^3}+\frac{28}{81}v_c^{\frac{1}{3}}\end{equation}
and
\begin{equation}p_c=\frac{3q_M^2}{v_c^4}+\frac{7}{54}\frac{1}{v_c^\frac{2}{3}}.\end{equation}
The equations (3.26) and (3.27) show that the critical temperature
$t_c$ and the critical pressure $p_c$ are sensitive to the
magnetic charge of the black hole $q_M$ just at small values of
the critical volume $v_c<<1$ but not for
its large amounts. \\
For positive temperatures $t\geq0$ (here  we exclude negative
temperatures because they are in usual way un-physical) one can
infer that $p-v$ diagram plotted by the equation (3.24) possess
inflection points under condition
\begin{equation}0\leq\Phi_E^2<1\end{equation} and these inflection points disappear for
electric potentials $\Phi_E^2\geq1.$ Because for (3.28)
coefficient of the term $\frac{1}{v_c^2}$ given by (3.24) remains
negative. Substituting
 (3.28) into the equation (3.25) we can obtain
admissible numerical values for the critical volume as
\begin{equation}0<\frac{6q_M^2}{v_c^2}-\frac{7}{27}v_c^\frac{4}{3}\leq1.\end{equation}
This shows upper and lower bounds of the critical volume $v_c$
which are become restricted by the magnetic charge $q_M.$ For
simplicity we choose equal sign in the above inequality identity
for which we will have
\begin{equation}q_M(\Phi_E=0)=\pm\frac{v_c}{18}\sqrt{14v_c^\frac{4}{3}+54}.\end{equation}
Substituting (3.30) into the relations (3.26) and (3.27) we obtain
\begin{equation}t_c=\frac{81+77v_c^\frac{4}{3}}{162v_c}\end{equation} and
\begin{equation}p_c=\frac{14v_c^2+27v_c^\frac{2}{3}}{54v_c^\frac{8}{3}}
\end{equation}
for zero electric potential $\Phi_E=0.$ Substituting (3.30) and
$\Phi=0$ into the equation (3.24) we obtain
\begin{equation}p=\frac{t}{v}+\frac{v_c^2}{v^4}\bigg(1+\frac{7}{27}v_c^\frac{4}{3}\bigg)
\bigg[\frac{1}{6}-\bigg(\frac{v}{v_c}\bigg)^2\bigg]
+\frac{7}{3}\frac{1}{v^{\frac{2}{3}}}\bigg[9+\bigg(\frac{v_c}{v}\bigg)^\frac{4}{3}\bigg].
\end{equation}
This shows that the for some positive constant temperatures $p-v$
diagrams will have inflection points only for
\begin{equation}\frac{v}{v_c}>\frac{1}{\sqrt{6}}\end{equation}
because of negativity sign of the term
$\big[\frac{1}{6}-\big(\frac{v}{v_c}\big)^2\big].$ Now we talk
about the enthalpy function (3.15) which under the condition
(3.30) reads
\begin{equation}m=\bigg(2+\frac{7}{27}v_c^\frac{4}{3}\bigg)v+\frac{v_c^2}{v}\bigg(1+\frac{7}{27}v_c^\frac{4}{3}\bigg)\bigg[\frac{1}{6}-
\bigg(\frac{v}{v_c}
\bigg)^2\bigg]-\frac{1}{v^\frac{7}{3}}+\frac{pv^3}{3}.\end{equation}
 Other
quantity which can be used to study thermal stability of black
holes is heat capacity at constant pressure defined by
\begin{equation}C_P=\bigg(\frac{\partial M}{\partial T}\bigg)_P\end{equation} which for $C_p<0$ the black hole is
unstable while for $C_p>0$ is stable. Applying (3.33) and (3.35)
the equation (3.36) reads
\begin{equation}c_p=\bigg(\frac{\partial m}{\partial t}\bigg)_p=\frac{W(v)}{O(v)}\end{equation}
in which $W(v)$ and $O(v)$ are obtained respectively as follows.
\begin{equation} W(v)=\bigg(\frac{\partial m}{\partial
v}\bigg)_p=2+pv^2+\frac{7}{3}\frac{1}{v^\frac{10}{3}}-\frac{v_c^2}{v^2}\bigg(1+
\frac{7}{27}v_c^\frac{4}{3}\bigg)\bigg[\frac{1}{6}+\bigg(\frac{v}{v_c
}\bigg)^2\bigg]
\end{equation} and
\begin{equation}O(v)=\bigg(\frac{\partial t}{\partial
v}\bigg)_p=p+\frac{v_c^2}{v^4}\bigg(1+\frac{7}{27}v_c^\frac{4}{3}\bigg)\bigg[\frac{1}{2}-\bigg(\frac{v}{v_c}
\bigg)^2\bigg]+\frac{7}{3}\frac{1}{v^\frac{2}{3}}\bigg[\bigg(\frac{v_c}{v}\bigg)^\frac{4}{3}-3\bigg]\end{equation}
with
\begin{equation}c_p=\frac{C_P}{8\pi r_q^2}.\end{equation}
Gibbs free energy of the quintessence dyonic AdS Black hole is
given  by (see Eq. (3.2) in ref. [10])
\begin{equation}G=M-TS-\Phi_EQ_E-\Phi_MQ_M=\mu N\end{equation}
where $S$ is entropy of the black hole which by according to the
Bekenstein-Hawking entropy formula is quarter of the black hole
horizon area [31] (see also [32]).
  \begin{equation}S=\int \frac{dM}{T}=\int_0^{r_h}\frac{dr_h}{T}\bigg(\frac{\partial M}{\partial r_h}\bigg)
  =\frac{4\pi r_h^2}{4}=\pi r_h^2=\frac{horizon~ area }{4}.\end{equation}
 $M$ is the black hole mass which is equal to the black hole
enthalpy energy. $T$ is the black hole Hawking temperature and
$\Phi_{E,M}$ and $Q_{E,M}$ are electromagnetic potential and
electromagnetic charge of the black hole. $N$ is number of micro
particles and $\mu$ is their mutual chemical potential.
Substituting (3.5), (3.8), (3.10) and (3.39) into the Gibbs free
energy (3.38) one can infer
  \begin{equation} G=\frac{(1-\Phi^2_E)r_h}{4}-\frac{1}{4}\frac{Q_M^2}{r_h}-\frac{2\pi r_h^3 P}{3}-\frac{(2+3\omega)r_q^{1+3\omega}}{4
  r_h^{3\omega}}
  \end{equation}
 which by substituting (3.13)  leads to a dimensionless form as follows.
\begin{equation}
g(p,v)=(1-\Phi_E^2)v-\frac{q^2_M}{v}-\frac{(2+3\omega)}{v^{3\omega}}-\frac{pv^3}{3}
\end{equation} where we  defined dimensionless Gibbs free energy
as follows.
\begin{equation}g=\frac{4G}{r_q}.\end{equation}
Substituting (3.23), (3.25) and (3.30) with $\Phi_E=0$ into the
black hole Gibbs free energy (3.44) we obtain

\begin{equation}g(p,v)=\frac{v_c^2}{v}\bigg(1+\frac{7}{27}v_c^\frac{4}{3}\bigg)\bigg[\bigg(\frac{v}{v_c}\bigg)^2-
\frac{1}{6}\bigg]+\frac{7}{27}v\bigg[9-\bigg(\frac{v_c}{v}
\bigg)^\frac{4}{3}\bigg]-\frac{pv^3}{3}
\end{equation}
Now we plot diagrams of state equation of the black hole $p(v,t)$,
its Gibbs free energy $g(p,v)$ versus the volume and the
temperature for different values of the pressure and then proceed
to interpretation of these diagrams. Diagrams of the thermodynamic
variables such as the pressure (3.33), the enthalpy (3.35), the
heat capacity (3.37), and Gibbs free energy (3.46) together with
all possible numerical values of the critical points of the
quintessence AdS dyonic black hole are plotted in figures 2 and
3.\\
Diagram of figure (2-a) shows all possible values for critical
volume obtained from the equation (3.25) by substituting $|q_M|>0$
and $0\leq\Phi_E^2<1.$ Setting $\Phi_E=0$ we plot all possible
values for $(q_M,v_c)$ by applying (3.30) in figure (2-b). The
figure (2-c) denotes variation of the critical compressibility of
the black hole $z_c=\frac{p_cv_c}{t_c}$ which can be calculated
from (3.31) and (3.32) versus the critical volume $v_c$. This
figure shows that for large $v_c$ the black hole system takes a
single phase, while for small $v_c$ the black hole system is made
from to subsystem (phase) with different compressibility factor
which they reach to a coexistence stable state with smallest value
of $v_c.$ Figure (2-d) describes variations of the critical
temperature versus the critical volume containing a local minimum
value. In other words by decreasing and increasing the critical
volume the critical temperature raises. We show absolutely
decreasing behavior of the critical pressure by raising the
critical volume in the figure (2-e). Applying (3.31) and  (3.32)
we plot possible values of the  critical pressure against the
possible values of the  critical temperature in the figure (2-f).
This shows monotonically raising behavior for $p_c$ when the
temperature increases $t_c$ for approximate values
$t_c\succeq2300$ but not for lesser values. For small values of
the critical temperature, $p_c$ increases rapidly. Diagram of the
equation (3.33) is plotted in figure (2-g) at constant pressure.
This figure shows a local maximum and a local minimum critical
temperature for small and large volume of the black holes
respectively. Looking to this figure one can infer that for
temperatures $3.5<t<5.8$ there is three unstable state for the
black holes under consideration. Minimum of the temperature
diagram  appears at $v\approx0.18$ and all black holes with
$v<0.18$ is unstable and so they are hotter. They incline to reach
a cooler larger stable state thermally with volume $\sim0.18$.
This means small/large black hole phase transition. Also all black
holes with $v>0.18$ incline to compress so that become smaller one
with stable volume $\sim0.18$. This is also means that unstable
large black holes exhibit with a large/small phase transition.
This phase transition can be interpreted in figure (2-h) at
constant temperature for a given pressure $-20<p<60$. This domain
is depended to some fixed temperatures which we used to plot the
diagram. One can extract more physics from behavior of the black
hole by looking the figure (2-i). In contrary the figure (1-a) for
the Van der Waals fluid which has a single crossing point the
figure (2-i) predicts a double crossing points. A single crossing
point at (1-a) shows a coexistence  of a fluid containing two
subsystems or phase (gas/fluid),while (2-i) predicts that the
black hole matter contains two coexistence state because of
existence two crossing point. This shows that the black hole can
have 4 subsystems or phase. Importance of the diagram (2-i) with
respect to other diagrams (2-g) and (2-h) is this: We saw that by
setting $t=constant$ and $p=constnat$ in the figures (2-g) and
(2-h) there is obtained three particular points with different
volumes which predict three unstable state of the black hole
matter. Physically these 3 unstable state can be considered as
gas/fluid/solid of phase the black hole material. They do not
predict forth plasma phase of the matter which can be coexistence
with other phases of the black hole matter.  In contrary the
figure (2-i) namely variation of the black hole Gibbs free energy
versus the temperature can be predict the plasma phase of the
black hole matter which can be occurs under the particular
conditions.

Diagrams of the figures (3-a) and (3-b) show the above situations
via variations of the heat capacity of the black hole at constant
pressure versus the volume and the temperature respectively. We
know that a thermodynamical system become unstable when its heat
capacity takes negative values and stable state for its positive
values. These diagrams show that this black hole will be stable
$C_p>0$ thermally just when its volume is not more larger or
smaller than the critical volume $v_c=0.1.$ There is small volume
$t_s<0.1$ and large volume $v_L>0.1$ where sign of the heat
capacity is changed. This is also predicts only a small/large
black hole phase transition which can be contain three phase of
the black hole matter. Diagram of the figure (3-b) shows a phase
of the black hole matter which by raising the temperature the heat
capacity has zero value $C_v\approx0$ while two other phases of
the black hole matter behave as $heat-giveier$ $(C_p<0)$ and
$heat-receiver$ $(C_p>0).$ There is a particular temperature
$t\to1.5$ where the heat capacity takes a degenerate state. Last
diagram (3-c) shows the enthalpy which has positive and negative
values for large and small black holes respectively. When we say
black hole is small
then is should be comparable with the critical volume $v_c=0.1$.\\
Last point which should be presented is this: All diagrams are
plotted easily except (g-t). This has more sensitive to initial
values. For instance with $p=30$ and $0<v<1$ the diagram shows the
black hole have two crossing points  but for  the number of these
crossing points can be reduced by changing the pressure.  Our
numerical calculations predicts they can be even disappeared by
changing the volume domain and the pressure. However as future
work, it's worth looking at finding more points of this type which
predict different phases of the black hole material. If there is
obtained more than 2 crossing point then one can result that the
black hole  exhibits with other  phases of the matter called as
the `Bose-Einstein  condensate` state and the `Fermionic
condensate` state which are fifth and sixth superfluid phase of
the matter respectively [34,35].

\section{Conclusion} In this work we studied quintessence dark
energy effects on the phase transition of the AdS dyonic black
holes. We determined regimes on the barotropic index of the
quintessence dark energy and electric potential and magnetic
charge of the black hole where the small/large black hole phase
transition is occurred. Also diagram of the p-v curves predicts
for large values of the thermodynamic volume of the black hole
phase transition can be reach to the well known Hawking-Page phase
transition where the black hole disappears and a vacuum AdS space
remains. Our work other phases of the black hole matter is
predicted such as $plasma,$ $Bose-Einstein$ and $Fermionic$
condensate in addition to three well known states called as
$solid-fluid-gas$ of the matter.  One of applicable work which we
can be investigate it in the future is study quintessence dark
energy effects on entanglement entropy of AdS dyonic black holes
in relation to the Maxwell equal area law which does not satisfied
for  AdS dyonic black holes  [28].
  \vskip 1cm
 \noindent
  {\bf References}\\
\begin{description}
\item[1.] R. Arnowitt, S. Deser, and C. W. Misner,~{\it {Canonical Variables for General Relativity}}, Phys. Rev. {\bf 117}, 1595, (1960).
\item[2.] R. Arnowitt, S. Deser, and C. W. Misner,~{\it {Coordinate Invariance and Energy Expressions in General Relativity}},
Phys. Rev. {\bf 122}, 997, (1961).
\item[3.] M. Henneaux and C. Teitelboim,~{\it {Asymptotically anti-de Sitter spaces}}, Commun. Math. Phys. {\bf 98}, 391, (1985).
\item[4.] S. Wang, S-Q. Wu, F. Xie and L. Dan,~{\it {The first law of thermodynamics of the (2+1)-dimensional
BTZ black holes and Kerr-de Sitter spacetimes}}, Chin. Phys. Lett.
{\bf 23}, 1096, (2006).
\item[5.] Y. Sekiwa,~{\it {Thermodynamics of de Sitter Black Holes: Thermal Cosmological Constant}},
Phys. Rev., D{\bf 73}, 084009, (2006).
\item[6.] K. Ball,~ {\it Volume ratios and a reverse isoperimetric inequality}, math/9201205 [math.MG]
\item[7.] M. Cvetic, G. Gibbons, D. Kubiznak, and C. Pope,~{\it {Black Hole Enthalpy and an Entropy Inequality
for the Thermodynamic Volume}}, Phys. Rev.D {\bf 84},  024037,
(2011).
\item[8.] B. P. Dolan,~{\it {Pressure and volume in the first law of black hole thermodynamics}},
 Class.Quant.Grav. {\bf 28}, 235017, (2011); gr-qc/1106.6260.
\item[9.] D. Kubizˇn´ak and R. B. Mann,~{\it { P − V criticality of charged AdS black holes}}, JHEP. {\bf 2012}, 033, (2012).
\item[10.] S. Dutta, A. Jain and R. Soni,~{\it {Dyonic Black Hole and Holography}}, JHEP {\bf 2013}, 60, (2013) hep-th/1310.1748.
\item[11.] X. X. Zeng and L. F. Li,~{\it {“Van der Waals phase transition in the framework of holography}}”, hep-th/1512.08855.
\item[12.] S. A. Hartnoll, C. P. Herzog and G. T. Horowitz, Phys. Rev. Lett. {\bf 101}, 031601, (2008), hep-th/0803.3295.
\item[13.] S. A. Hartnoll, C. P. Herzog and G. T. Horowitz, JHEP {\bf 0812}, 015, (2008), hep-th/0810.1563.
\item[14.] S. A. Hartnoll and P. Kovtun, Phys. Rev. D {\bf 76}, 066001, (2007), hep-th/0704.1160.
\item[15.] M. M. Caldarelli, O. J. C. Dias and D. Klemm, JHEP {\bf 0903}, 025, (2009), hep-th/0812.0801.
\item[16.] S. A. Hartnoll, P. K. Kovtun, M. Muller and S. Sachdev, Phys. Rev. B {\bf 76}, 144502, (2007),cond-mat.str-el/0706.3215.
\item[17.] N. A. Bachall, J. P. Ostriker, S. Perlmutter and P. J. Steinhardt, {\it {The cosmic triangle: Revealing the state of the universe}},
Science {\bf 284}, 1481, (1999).
\item[18.] S. J. Perlmutter et al, {\it {Measurements of Omega and Lambda from 42 High-Redshift Supernovae}}, Astrophys. J. {\bf 517}, 565, (1999).
\item[19.] V. Sahni and A. A. Starobinsky, {\it The case for a positive cosmological Lambda-term}, Int. J. Mod. Phys. D {\bf 9}, 373, (2000).
\item[20.] Shinji Tsujikawa, {\it {Quintessence: A Review}}, Class. Quant. Grav. {\bf 30}, 214003, (2013).
\item[21.] L. H. Ford, {\it {Cosmological-constant damping by unstable scalar fields}}, Phys. Rev. {\bf D 35}, 2339, (1987).
\item[22.] Y. Fujii, {\it {Origin of the gravitational constant and particle masses in a scale invariant scalar-tensor theory}}, Phys. Rev.
 {\bf D 26}, 2580, (1982).
\item[23.] V. V. Kiselev, {\it {Quintessence and black holes}}, Class. Quant. Grav. {\bf 20}, 1187, (2003); gr-qc/0210040.
\item[24.] Y. Zhang, Y. X. Gui and F. L. Li, {\it {Quasinormal modes of a Schwarzschild black hole surrounded by
 quintessence: Electromagnetic perturbations}}, Gen. Rel. Grav. {\bf 39}, 1003, (2007).
\item[25.] N. Varghese and V. C. Kuriakose, {\it {Massive Charged Scalar Quasinormal Modes of Reissner-Nordstrom Black Hole
 Surrounded by Quintessence}}, Gen. Rel. Grav. {\bf 41}, 1249, (2009).
\item[26.] S. Chen, Q. Pan and J. Jing, {\it {Holographic superconductors in quintessence AdS black hole spacetime}},
Class. Quant. Grav. {\bf 30}, 145001, (2013).
\item[27.]    R. H. Swendsen, {\it An Introduction to Statistical Mechanics and
Thermodynamics}; (Oxford University Press 2012.)
\item[28] P. H. Nguyen, {\it An equal area law for holographic entanglement entropy of the AdS-RN black
hole}, JHEP, {\bf 12}, 139, (2015); hep-th/1508.01955.
\item[29.] G. Q. Li, {\it `Effects of dark energy onP-Vcriticality of chargedAdS black
holes`}, Phys. Lett. {\bf B06}, 260 (2014); gr-qc/1407.0011
\item[30.]  H. Liu and X.H. Meng, {\it Effects of dark energy on the efficiency of charged AdS black holes as heat
engine}     Eur. Phys. J. C {\bf 77}, 556, (2017);
hep-th/1704.04363v4.
\item[31.] J. D Bekenstein, {\it Black Holes and Entropy} Phys.
Rev. D{\bf7}, 2333, (1973).
\item[32.] E. Papantonopoulos {\it Physics of Black holes, `Lecture notes in physics
769`; Springer-Verlag Berlin Heidelberg 2009.  }
\item[33.] S. W. Hawking and D. N. Page,
{\it Thermodynamics Of Black Holes In Anti-De Sitter Space} ,
Commun. Math. Phys. {\bf 87}, 577, (1983).
\item[34.] B. DeMarco; J. Bohn and E. Cornell
{\it Pioneer of ultracold quantum physics} , Nature {\bf 538},
318, (2006).
\item[35.] C. A. Regal; M. Greiner and D. S Jin
{\it Observation of Resonance Condensation of Fermionic Atom
Pairs}, Phys. Rev. Lett{\bf  92}, 040403, (2004); cond-mat/0401554
.

\end{description}
\begin{figure}[ht] \centering
\hspace{1mm} \subfigure[{}]{\label{1}
\includegraphics[width=.3\textwidth]{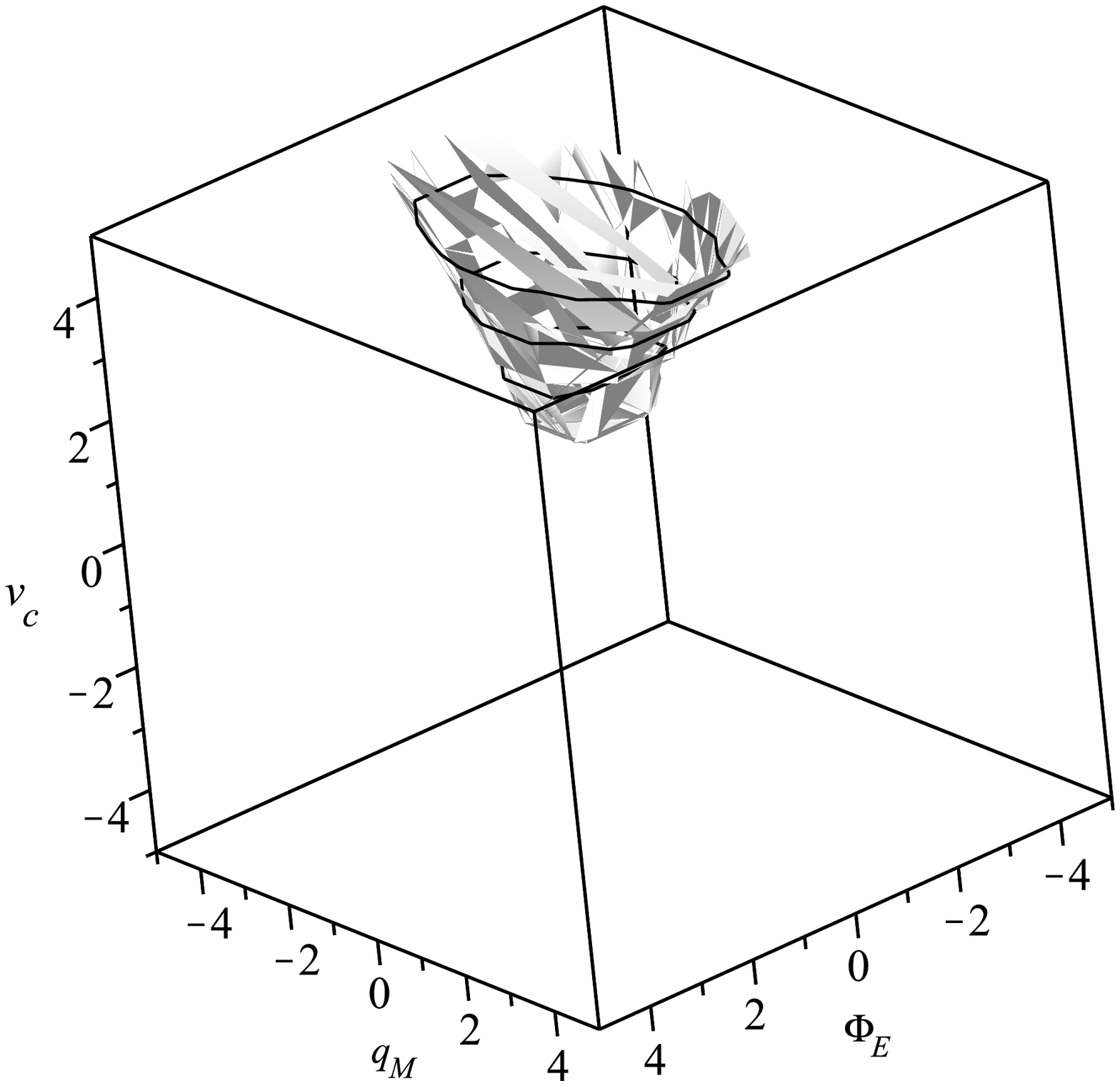}}
\hspace{1mm} \subfigure[{}]{\label{1}
\includegraphics[width=.3\textwidth]{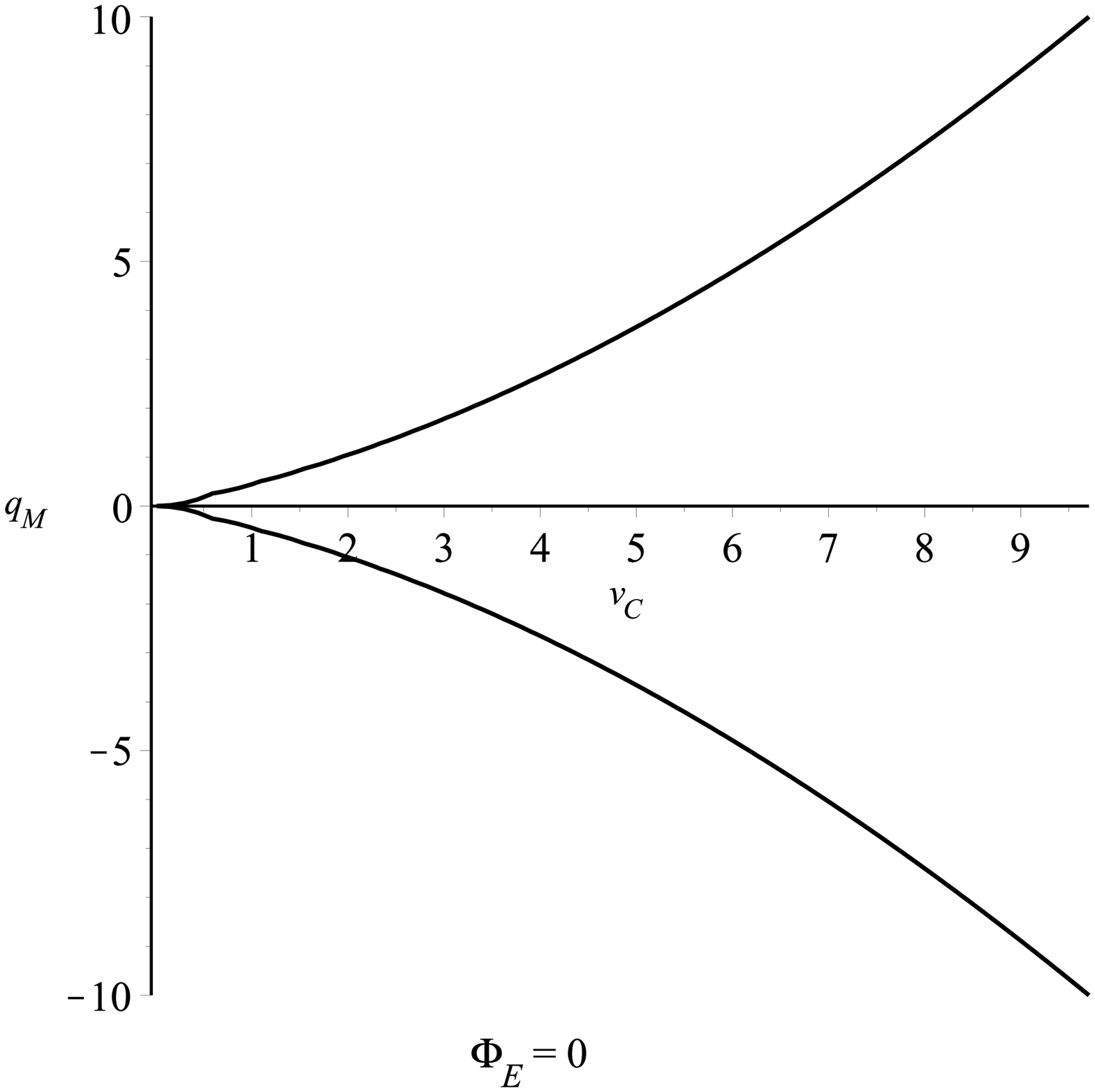}}
\hspace{1mm} \subfigure[{}]{\label{1}
\includegraphics[width=.3\textwidth]{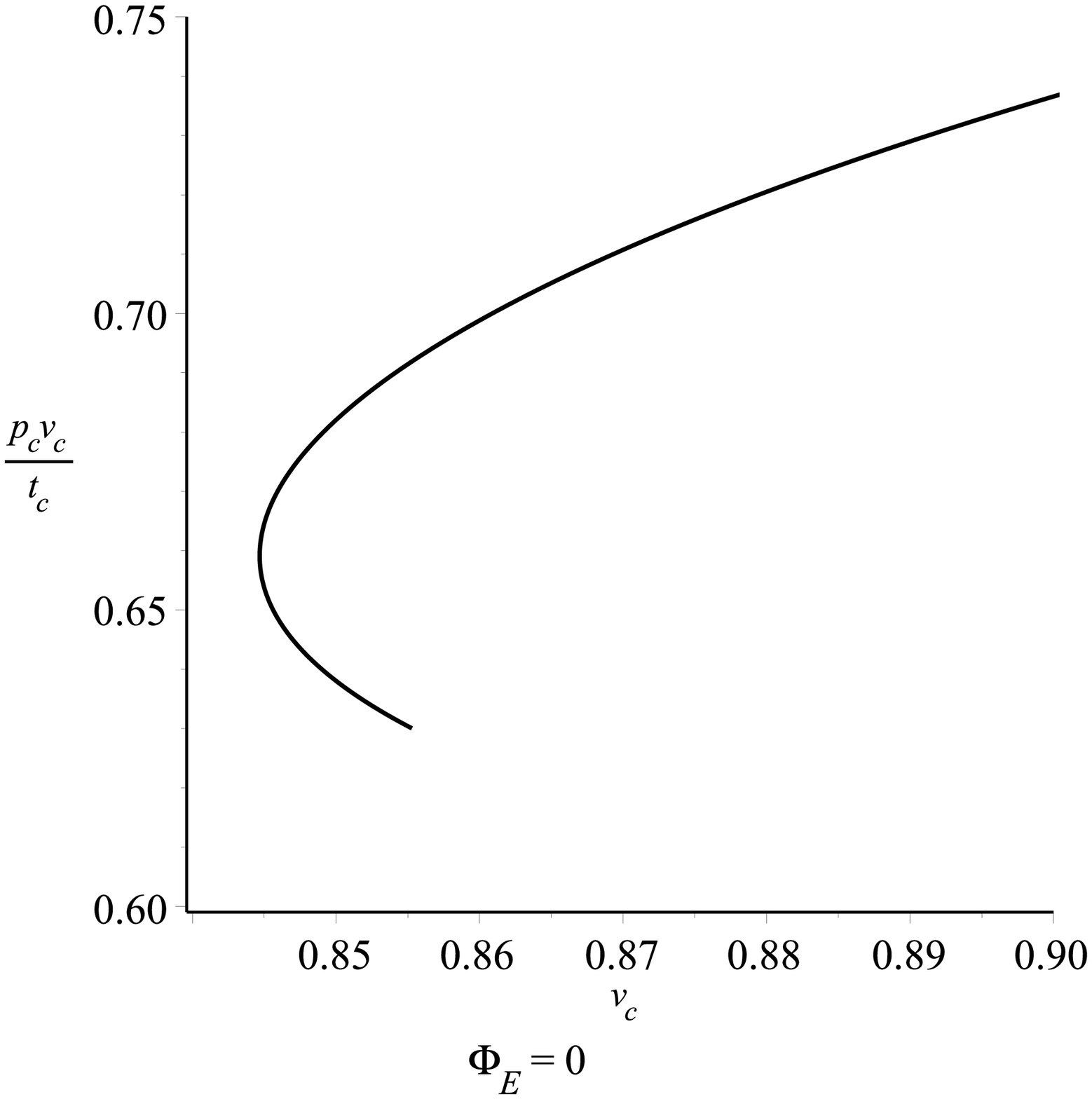}}
\hspace{1mm} \subfigure[{}]{\label{1}
\includegraphics[width=.3\textwidth]{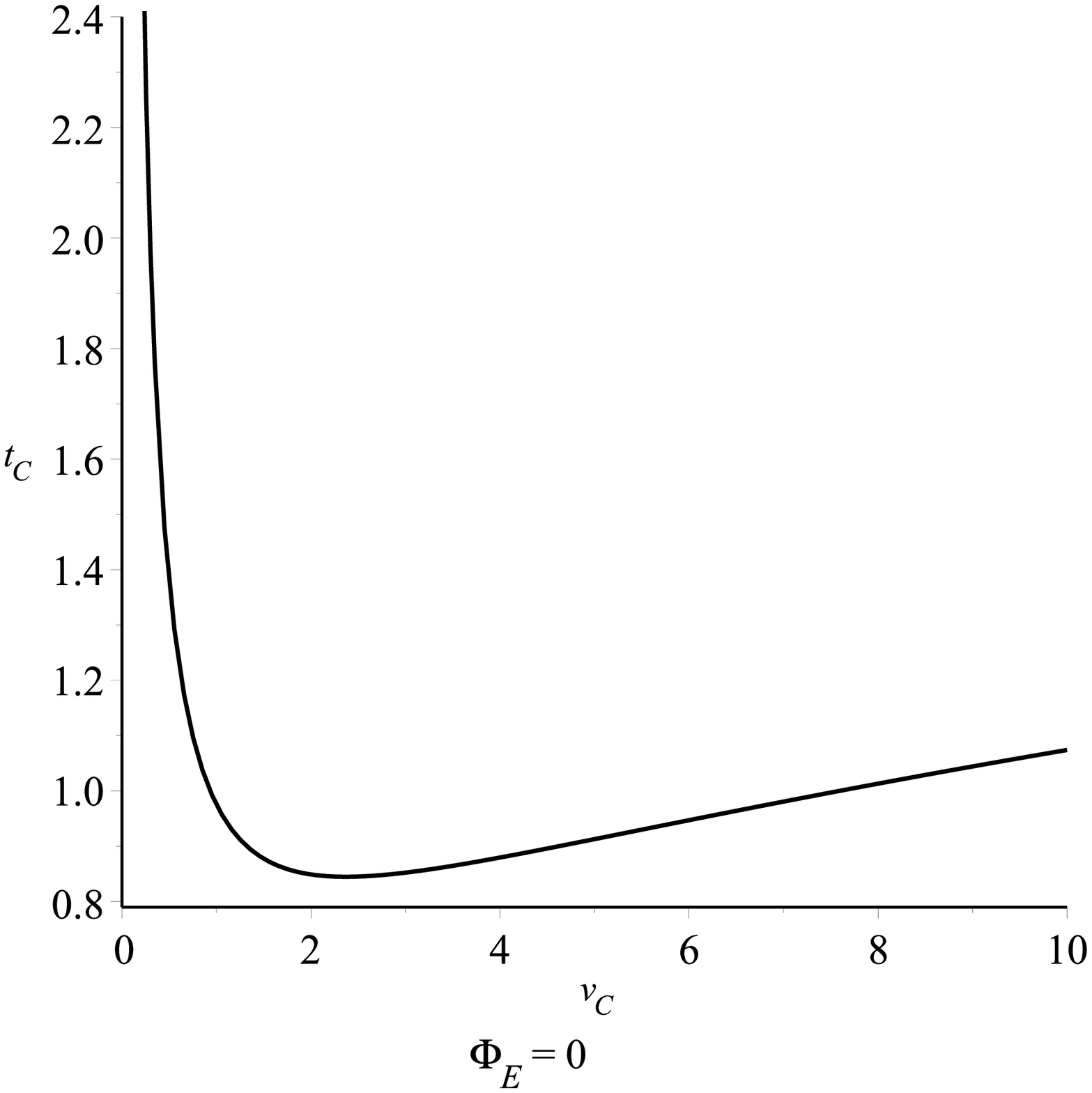}}
\hspace{1mm} \subfigure[{}]{\label{1}
\includegraphics[width=.3\textwidth]{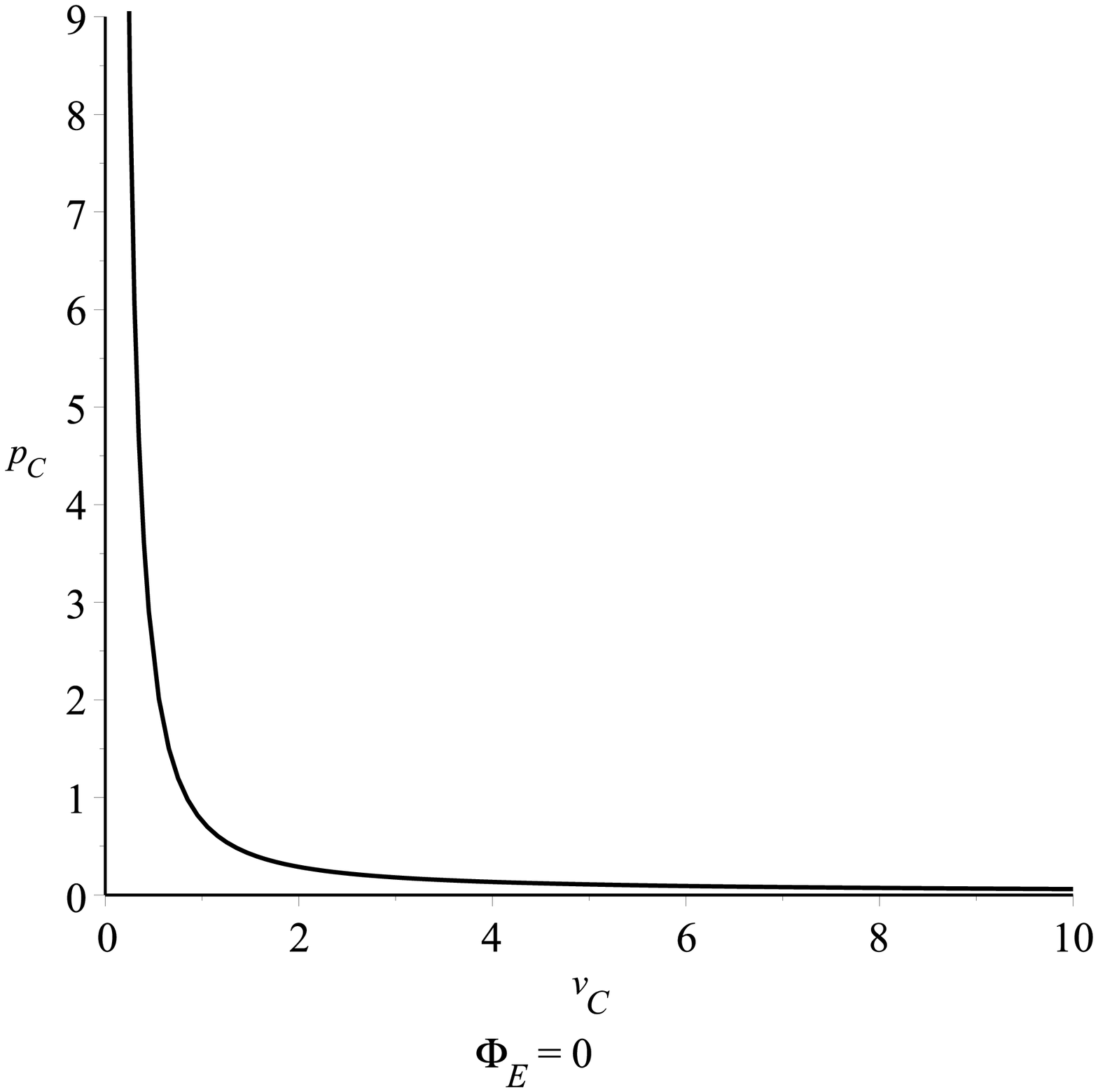}}
\hspace{1mm} \subfigure[{}]{\label{1}
\includegraphics[width=.3\textwidth]{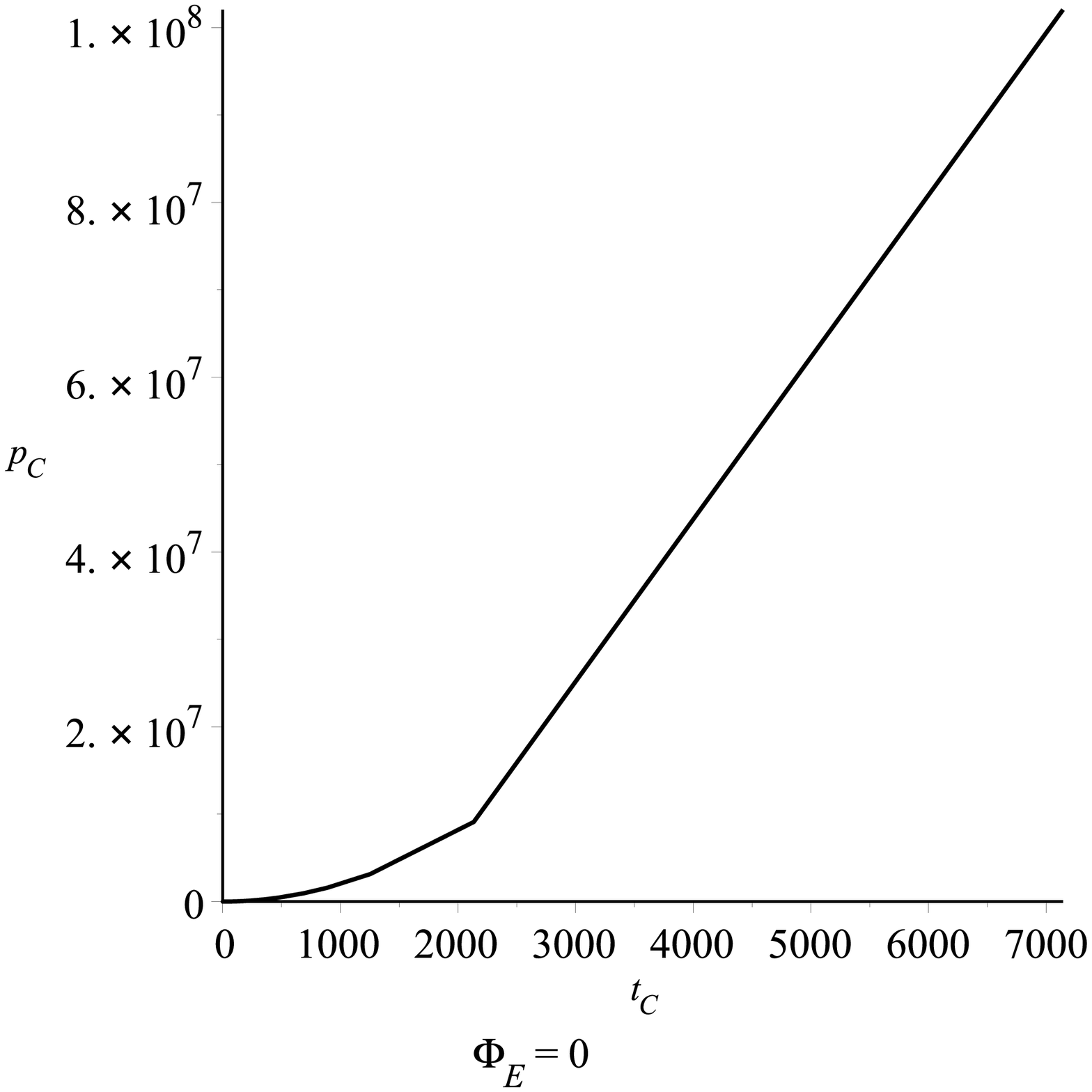}}
\hspace{1mm} \subfigure[{}]{\label{1}
\includegraphics[width=.3\textwidth]{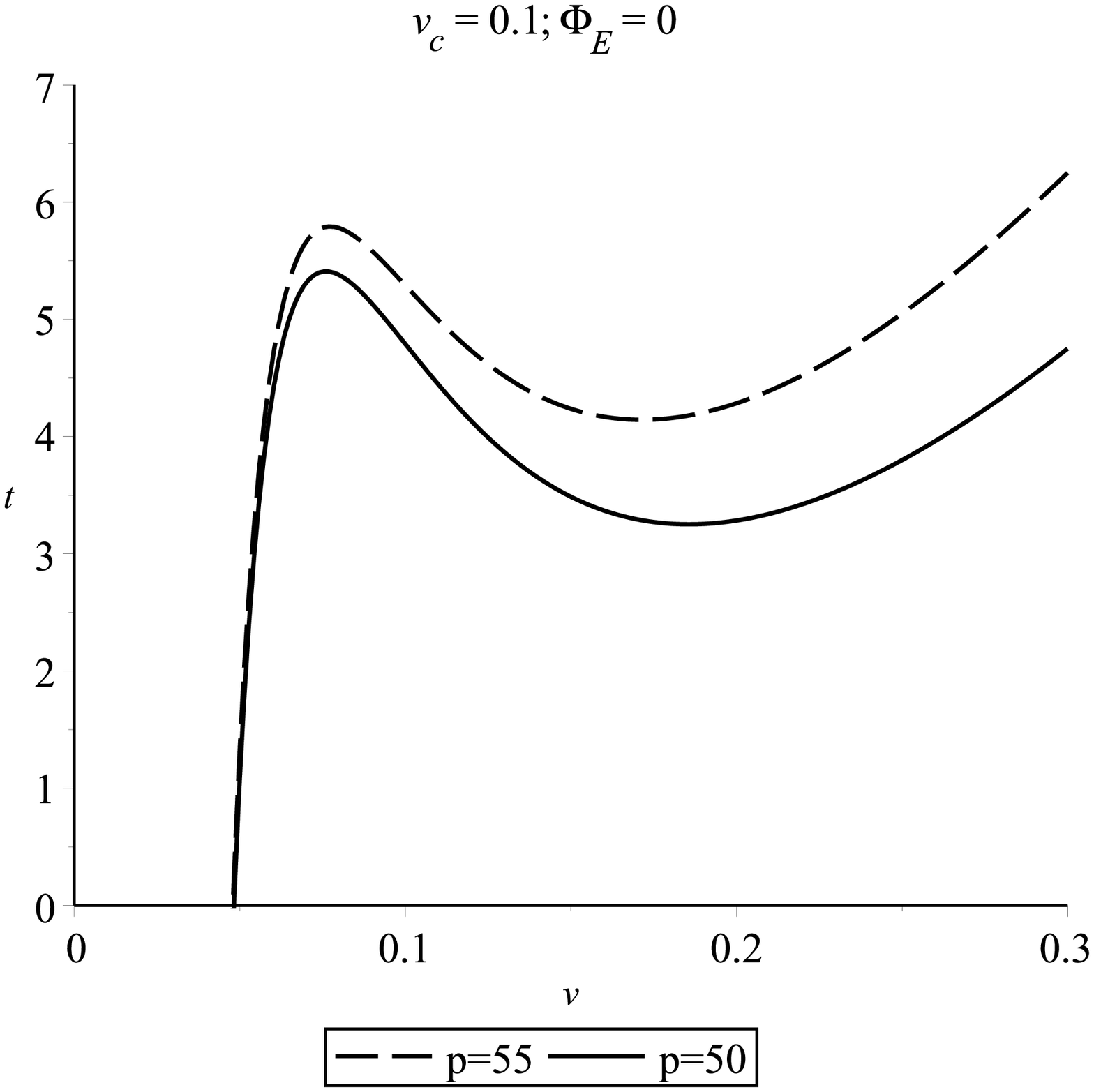}}
\hspace{1mm} \subfigure[{}]{\label{1}
\includegraphics[width=.3\textwidth]{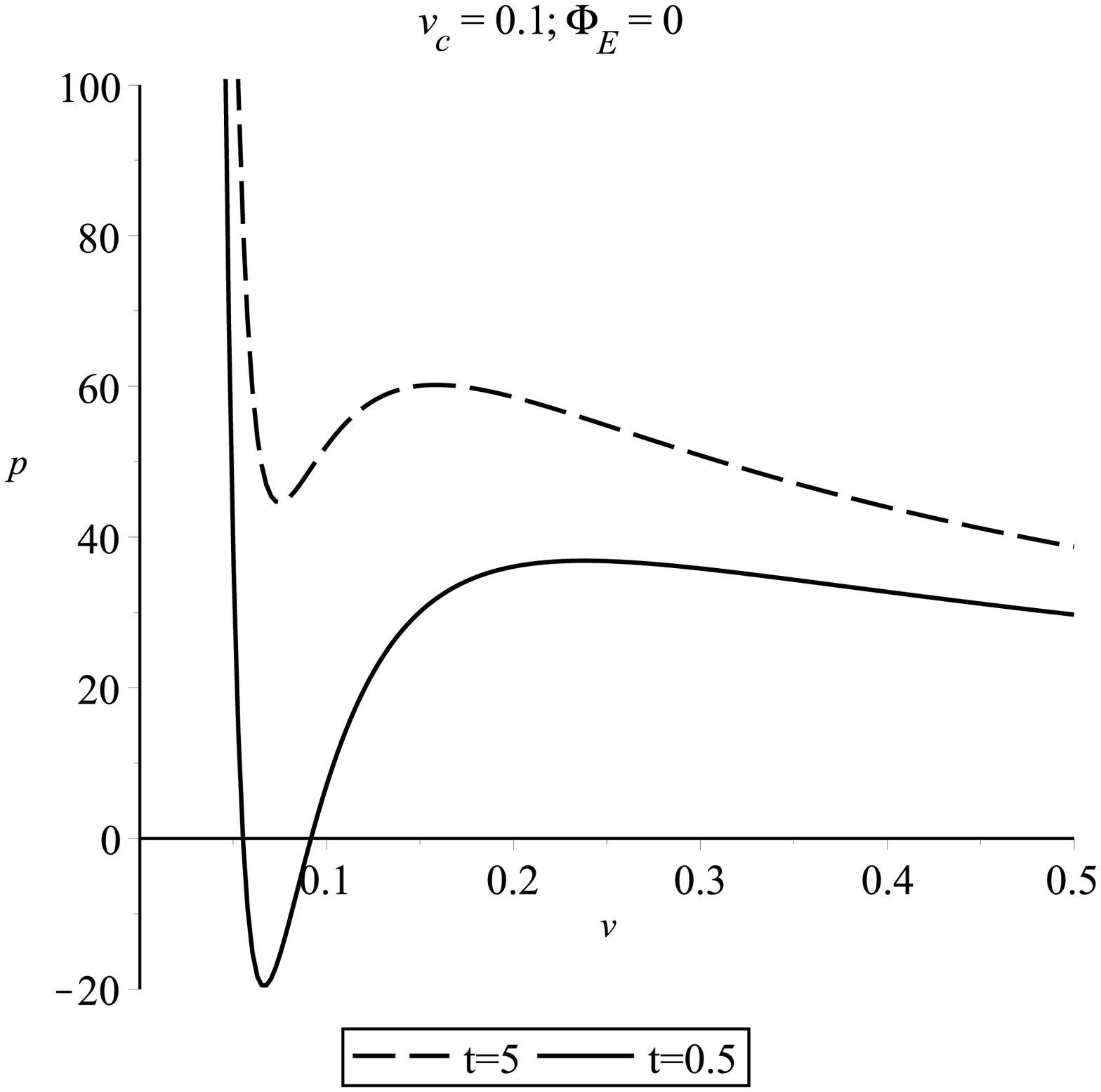}}
\hspace{1mm} \subfigure[{}]{\label{1}
\includegraphics[width=.3\textwidth]{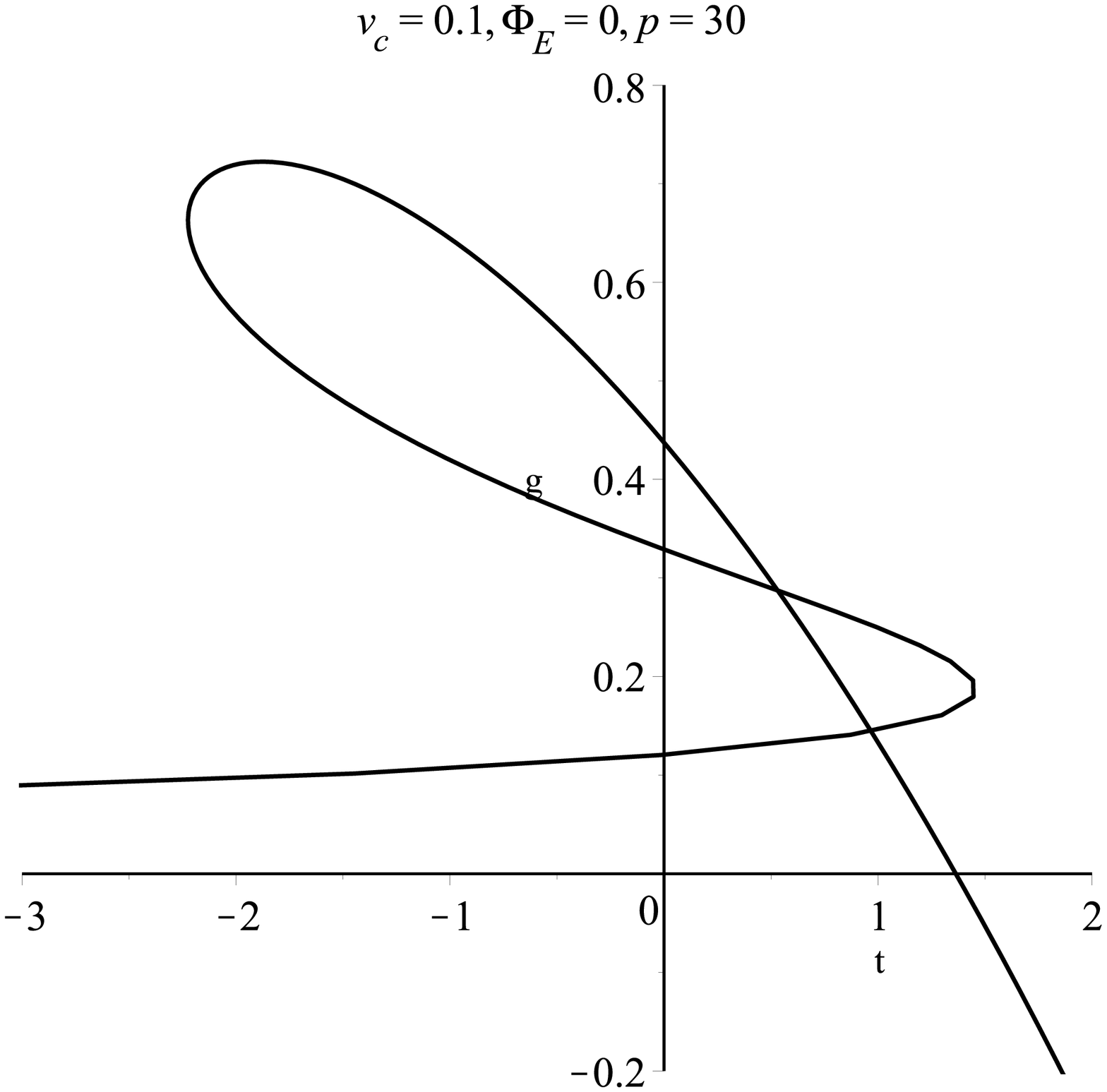}}
\caption{Diagram of critical points and small/large phase
transition of the quintessence AdS Dyonic black hole} \label{l}
\end{figure}

\begin{figure}[ht] \centering\hspace{1mm} \subfigure[{}]{\label{1}
\includegraphics[width=.3\textwidth]{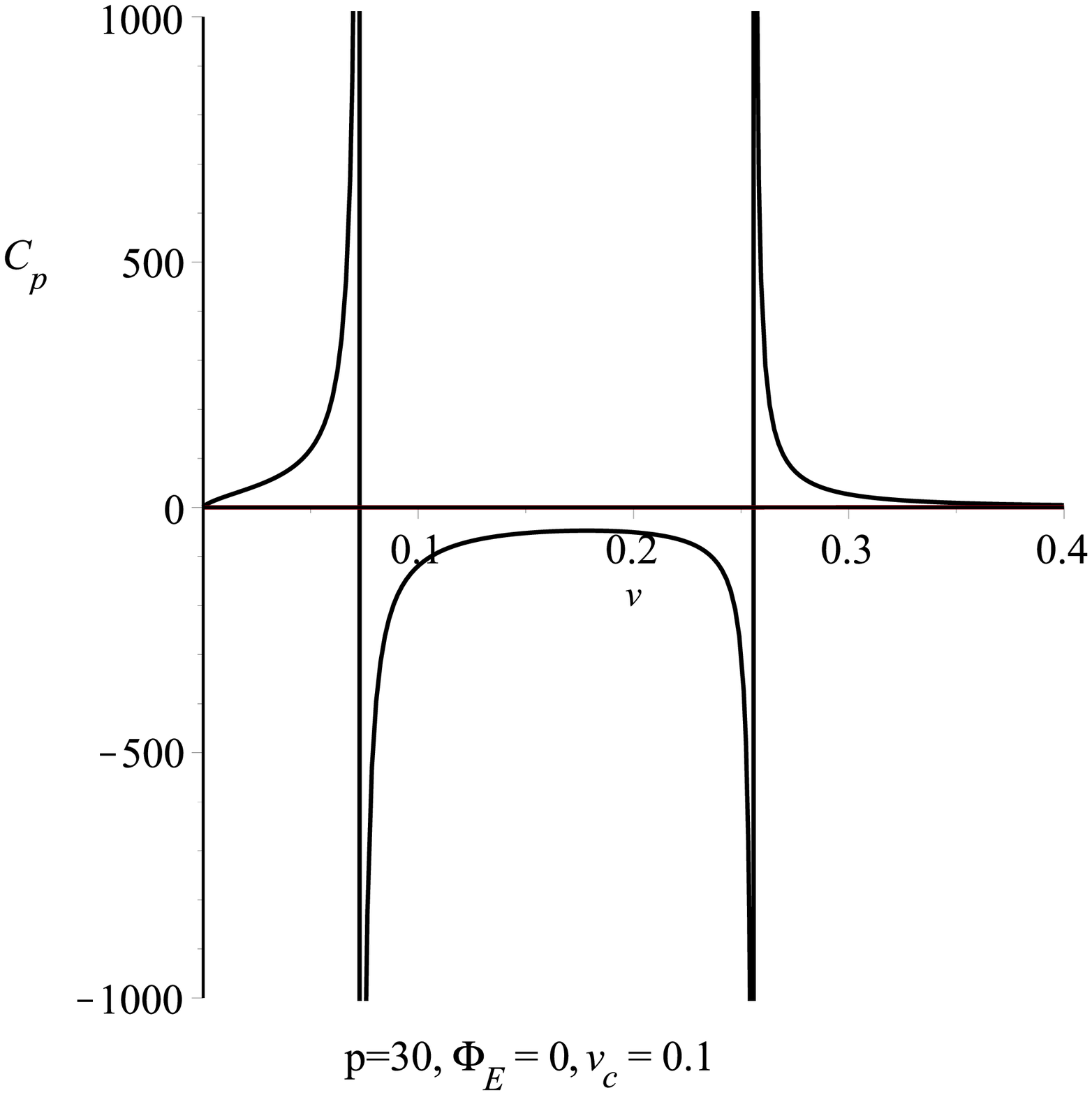}}
\hspace{1mm} \subfigure[{}]{\label{1}
\includegraphics[width=.3\textwidth]{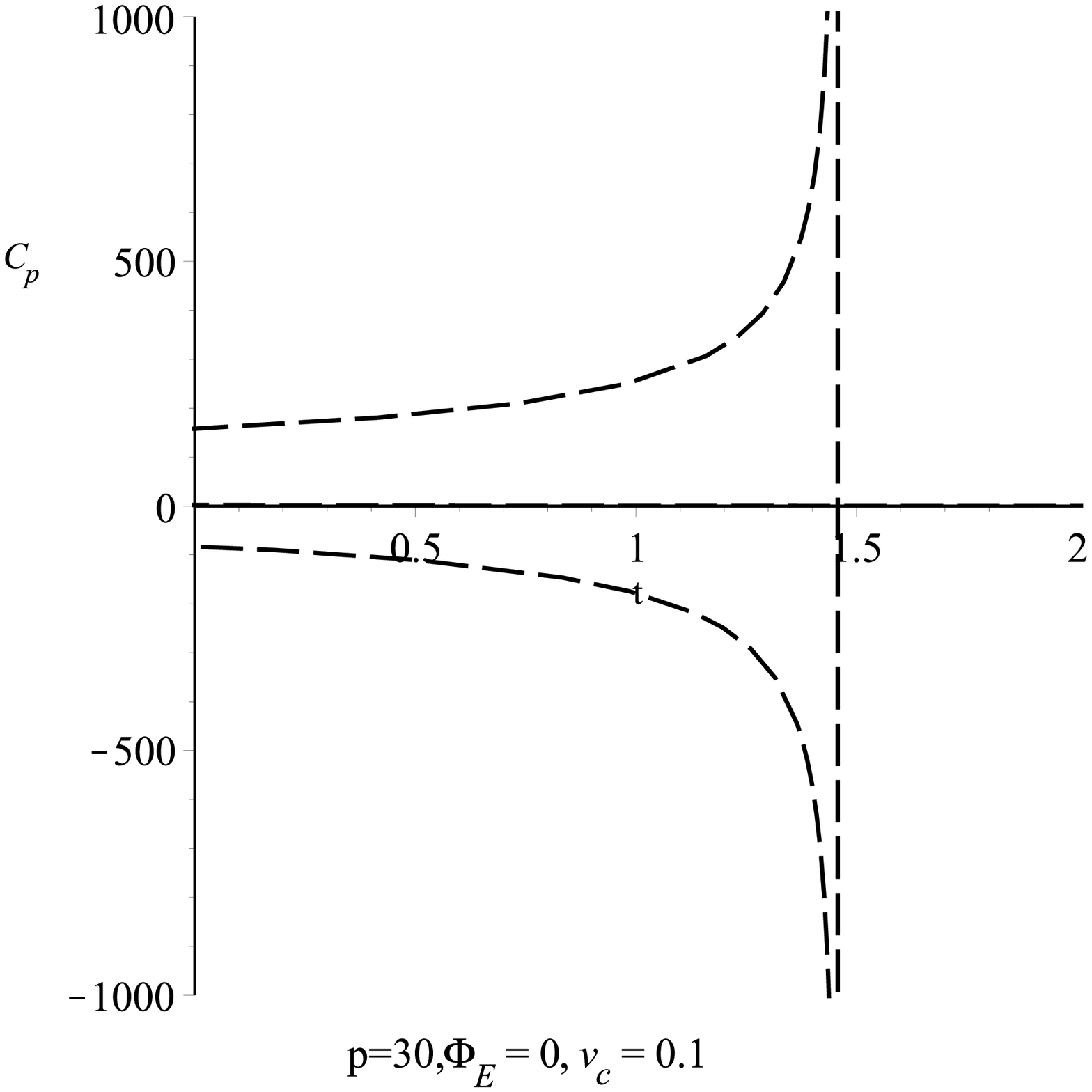}}
\hspace{1mm} \subfigure[{}]{\label{1}
\includegraphics[width=.3\textwidth]{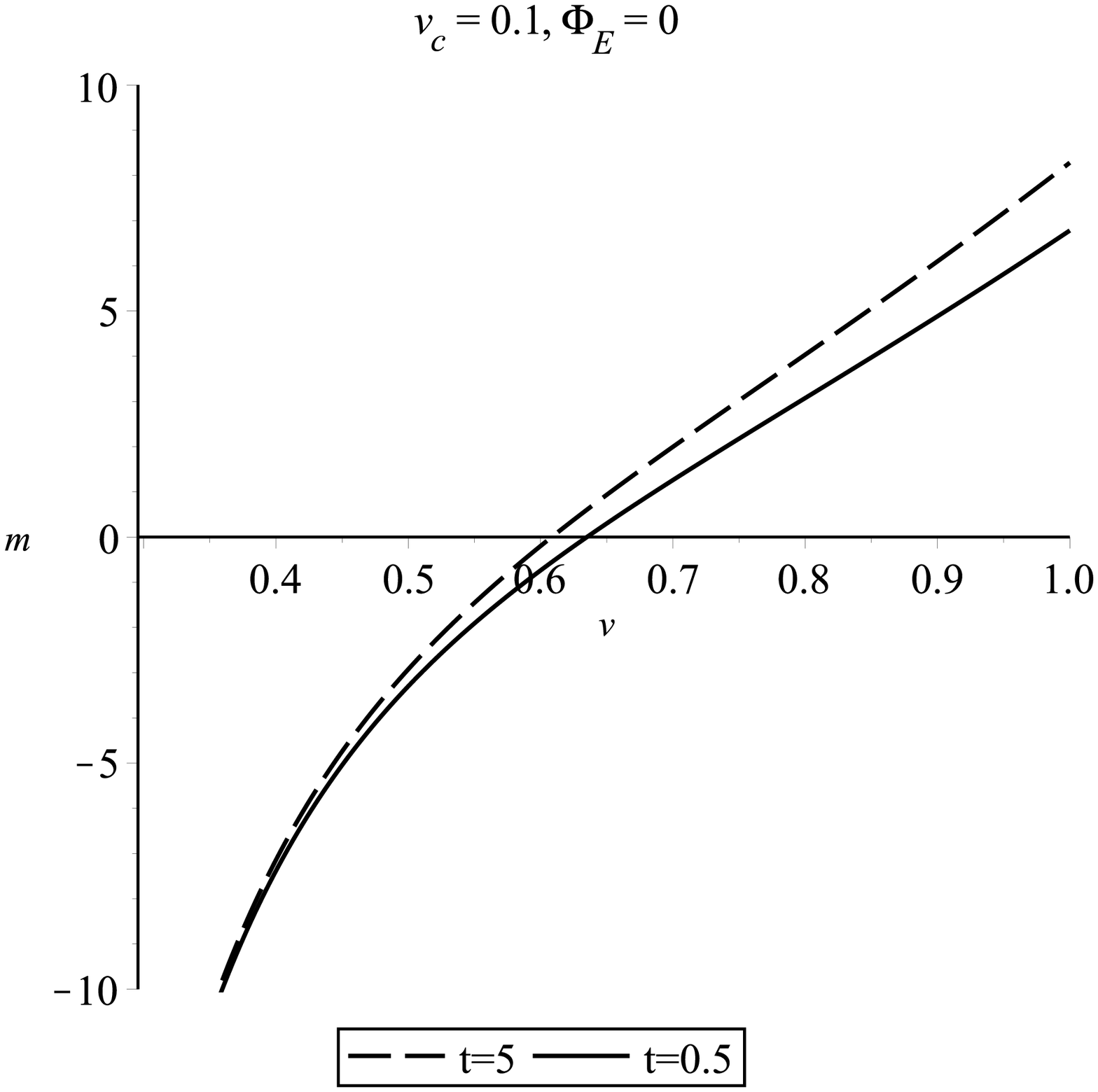}}
\caption{Diagrams of the heat capacity are plotted versus the
volume and the temperature at (a) and (b) respectively and the
black hole enthalpy is plotted versus the volume at (c).
}\label{l}
\end{figure}

\end{document}